\documentclass[aps, prb, twocolumn,floatfix, showpacs, 10pt, superscriptaddress]{revtex4-1}
\usepackage{amsmath,amsfonts,amsbsy,amssymb,mathrsfs,bm,braket,wasysym}
\usepackage{graphicx}
\usepackage{textcomp}
\usepackage[caption=false,singlelinecheck=false]{subfig}
\usepackage{xcolor}
\usepackage{ulem}
\usepackage{gensymb}
\usepackage[super]{nth}
\usepackage[colorlinks,linkcolor=blue,citecolor=cyan,filecolor=magenta,urlcolor=cyan]{hyperref}

\allowdisplaybreaks

\begin{document}
\title{Emergent chiral spin ordering and anomalous Hall effect of kagome lattice at 1/3 filling}

\author{Hee Seung Kim}
\affiliation{Korea Advanced Institute of Science and  Technology, Daejeon, South Korea}
\author{Archana Mishra}
\affiliation{Korea Advanced Institute of Science and  Technology, Daejeon, South Korea}
\affiliation{International Research Centre MagTop, Institute of Physics, Polish Academy of Sciences, Aleja Lotnik´ow 32/46, PL-02668 Warsaw, Poland}
\author{SungBin Lee}
\affiliation{Korea Advanced Institute of Science and  Technology, Daejeon, South Korea}

\date{\today}
\begin{abstract}
The study of electronic and magnetic properties of kagome lattice has been an active research area searching for topological phases of matters. In particular, the kagome system with transition metal stannides and etc exhibit interesting anomalous Hall effects driven by ferromagnetic or non-collinear magnetic ordering. In this paper, motivated by these pioneer works, we study strongly correlated spin-orbit coupled electrons in kagome lattice at 1/3 filling. Using both Hartree-Fock approach and effective model analysis, we report quantum phase transitions accompanied with distinct charge and magnetic ordered phases. Especially, for strongly interacting limit, we discover new types of \textit{1(2)-pinned metallic} states which are understood by effective localized electron models. Furthermore, when spin-orbit coupling is present, it turns out that such \textit{pinned metallic} states open a wide region of Chern insulating phase with chiral spin ordering. Thus, quantum anomalous Hall effect is expected with emergent scalar spin chirality. Our theory provides a theoretical platform of strongly interacting kagome metal which is applicable to transition metal stannides. 
\end{abstract}
\maketitle

The kagome lattice has received a great attention to study band topology due to similar yet distinct band structure compared to the honeycomb lattice for non-interacting case\cite{PhysRevLett.106.236802,PhysRevB.80.113102,PhysRevB.79.035323,doi:10.1063/1.4739724,Mazin2014}. Two Dirac cones exist similar to honeycomb lattice, which are protected by time reversal, parity, and $C_3$ rotational symmetry. Thus, the inclusion of spin-orbit coupling (SOC) opens a non-trivial band gap which induces topological insulator (TI). In addition to Dirac cones, there is another unique property, the flat band, which indicates localized wave function due to destructive interference of wave function caused by geometrical frustration\cite{Liu_2014_flat,PhysRevB.78.125104}. Since SOC also induces topologically non-trivial flat band, the kagome lattice becomes a strong candidate for realizing fractional quantum Hall effect\cite{PhysRevLett.106.236804,PhysRevLett.106.236802}. Therefore, distinctive band structure of the kagome lattice is expected to give fruitful topological properties.

Considering electron correlations, on the other hand, the Hubbard model on the kagome lattice has been exactly solved based on graph-theory for filling $n\! \geq \! 5/6$.  At zero temperature, it has been shown that ferromagnetic ordering is stabilized at $5/6\!\leq \!n\!<\!11/12$, whereas, the system becomes paramagnetic at $n \geq \! 11/12$\cite{Mielke_1992}. Furthermore, ferromagnetism of kagome lattice model has been also studied exactly at 1/6 filling as an extension of earlier work with sufficiently large interaction strength\cite{PhysRevLett.100.136404}. Although for other fillings it is not exactly solvable, numerical studies show exotic spin and charge density wave phases\cite{PhysRevB.82.075125}. In particular, at 1/3 filling with intermediate interaction strength, unique magnetic ordered phase, so called the \textit{pinned metal droplet} phase, has been discussed that breaks lattice translation symmetry spontaneously\cite{PhysRevB.89.155141}. This phase, a consequence of unique lattice geometry of kagome lattice, can be understood by spin polarized electrons locally form a metal on hexagons. 

In strongly interacting limit, however, it has not been rigorously studied how such locally metallic phase changes to other phases with possible phase transitions. Furthermore, though the effect of SOC and electron correlation on the kagome lattice have been well studied independently, very few works are present about the interplay between SOC and electron correlation\cite{Liu_2014}. Thus, given that it's unique lattice geometry plays a crucial role to stabilize \textit{pinned metal droplet} phase even in the intermediate interaction strength, one may expect more exotic phases with strong interaction limit, interplay of SOC and kagome lattice geometry.
  
From the experimental point of view, it has been known that various transition metal stannides with a chemical formula $\text{A}_{x}\text{Sn}_{y}$ type (A = Mn, Fe, or Co; $x$:$y$ = 3:1, 3:2 or 1:1) form quasi two dimensional metallic kagome layer structures\cite{kang2019dirac,GIEFERS2006132}. These materials are the apt place to study the interplay of SOC and interaction. Especially,  $\text{Fe}_{3}\text{Sn}_{2}$ is spotlighted because of its ferromagnetic or non-collinear magnetic ordering with quantum anomalous Hall effect\cite{Fenner_2009,Ye2018,Kida_2011,kang2019dirac}. In this context, we are motivated to study underlying physics of quantum anomalous Hall effect and magnetism of $\text{Fe}_{3}\text{Sn}_{2}$ as a consequence of unique lattice geometry, strong interaction and SOC and their applications to other transition metal stannides. 

In this paper, we consider the Hubbard model with intrinsic SOC in the kagome lattice at 1/3 filling where the Fermi level is located at the Dirac point. We explicitly show the phase diagram as functions of SOC and interaction strength parameters, based on both Hartree-Fock approximation and effective model analysis. We discover new magnetic metallic phases in strong interaction limit, \textit{1-(2-) pinned metallic} phases where spin polarized electrons form coupled conducting chains and part of (1- and 2-) electrons are being localized (pinnned) with opposite spin. In the presence of SOC, the system further develops spin canting and Chern insulating phase is uniquely stabilized with emergent scalar spin chirality. We also discuss the phase transition to Chern ferromagnetic insulating phase in the strong interaction limit.

\begin{figure}
	\subfloat[]{\includegraphics[width=0.14\textwidth]{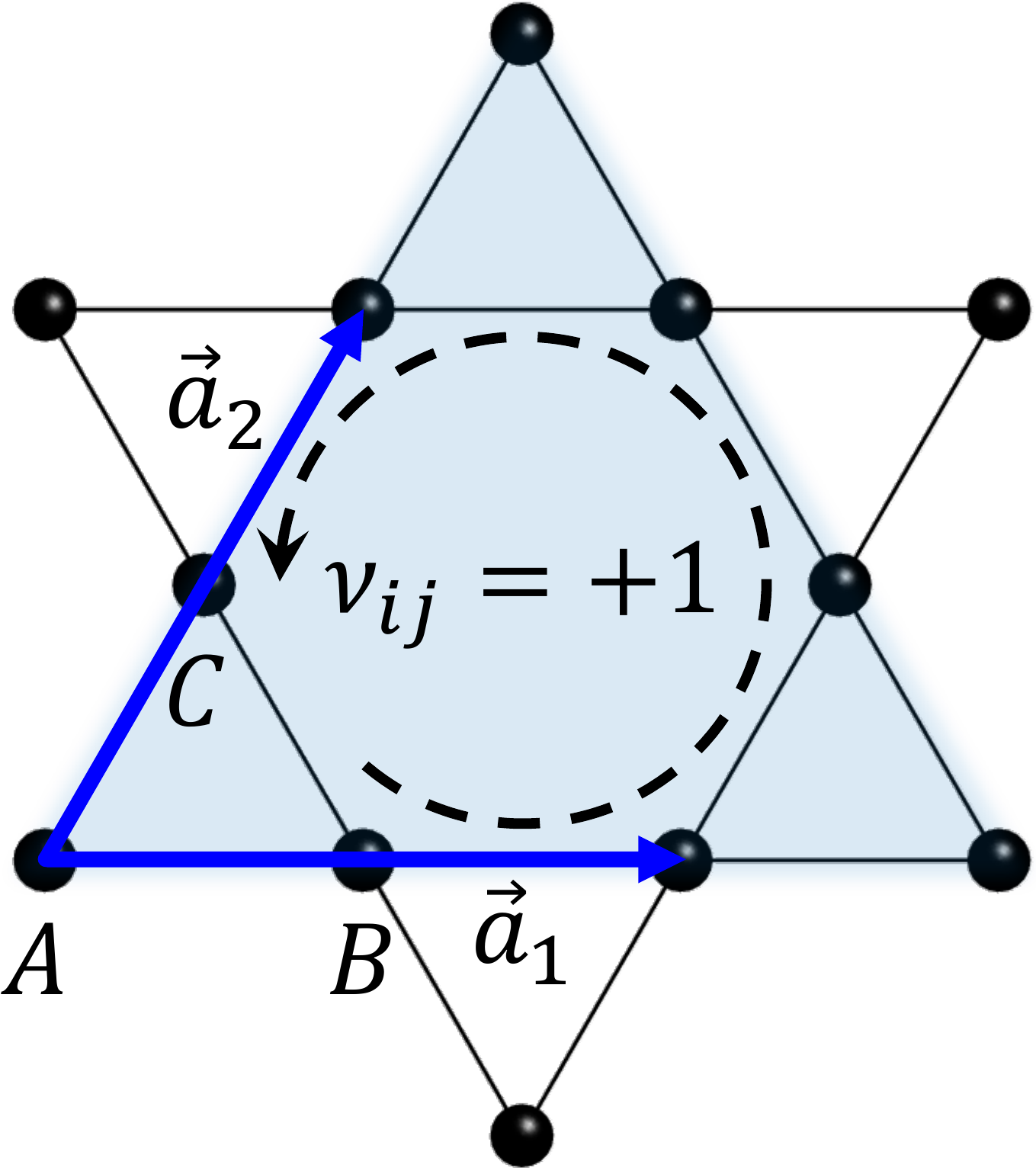}\label{fig:kagomelattice}} ~
	\subfloat[]{\includegraphics[width=0.14\textwidth]{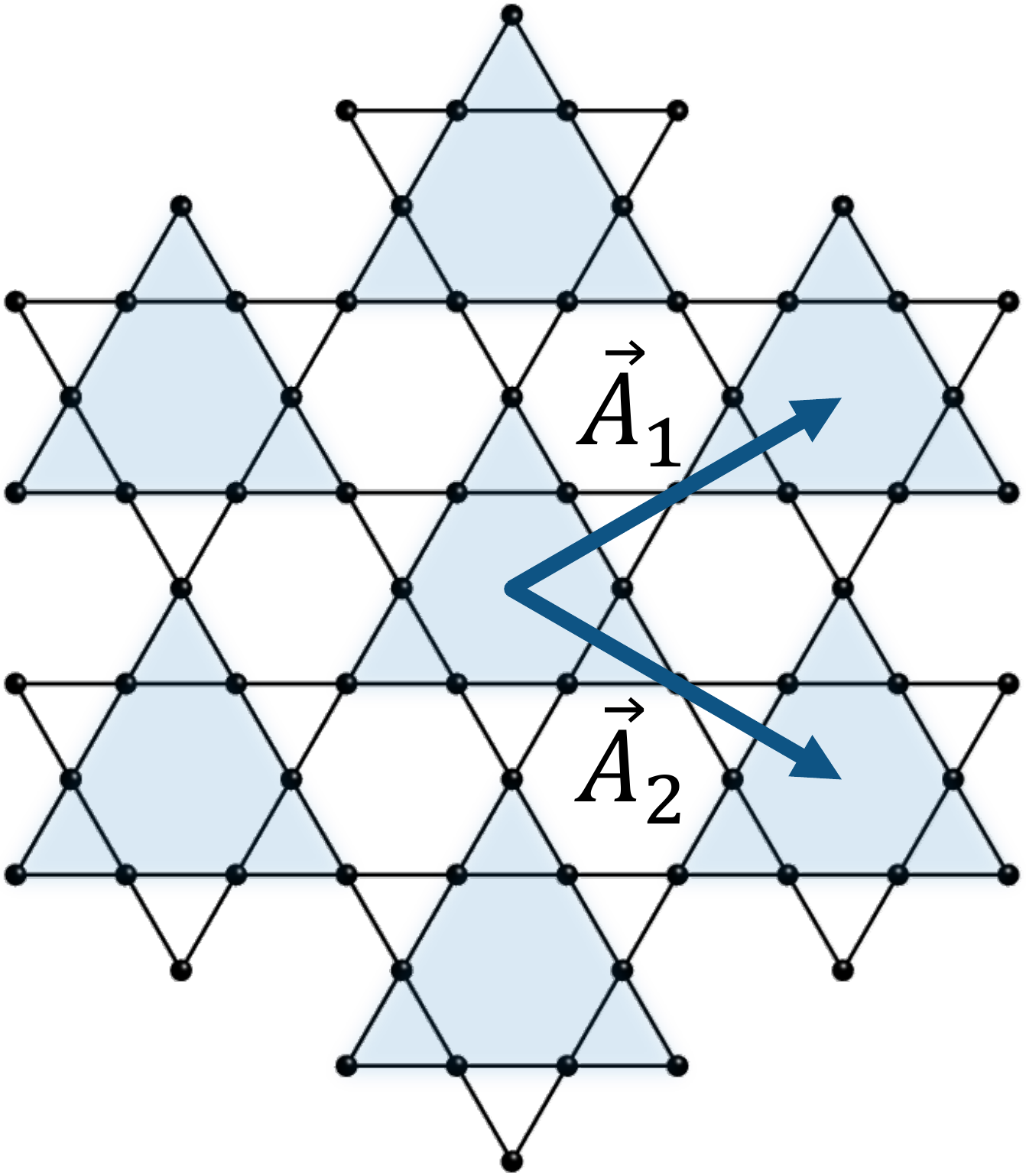}\label{fig:enlarged_kagome}} ~
	\subfloat[]{\includegraphics[width=0.14\textwidth]{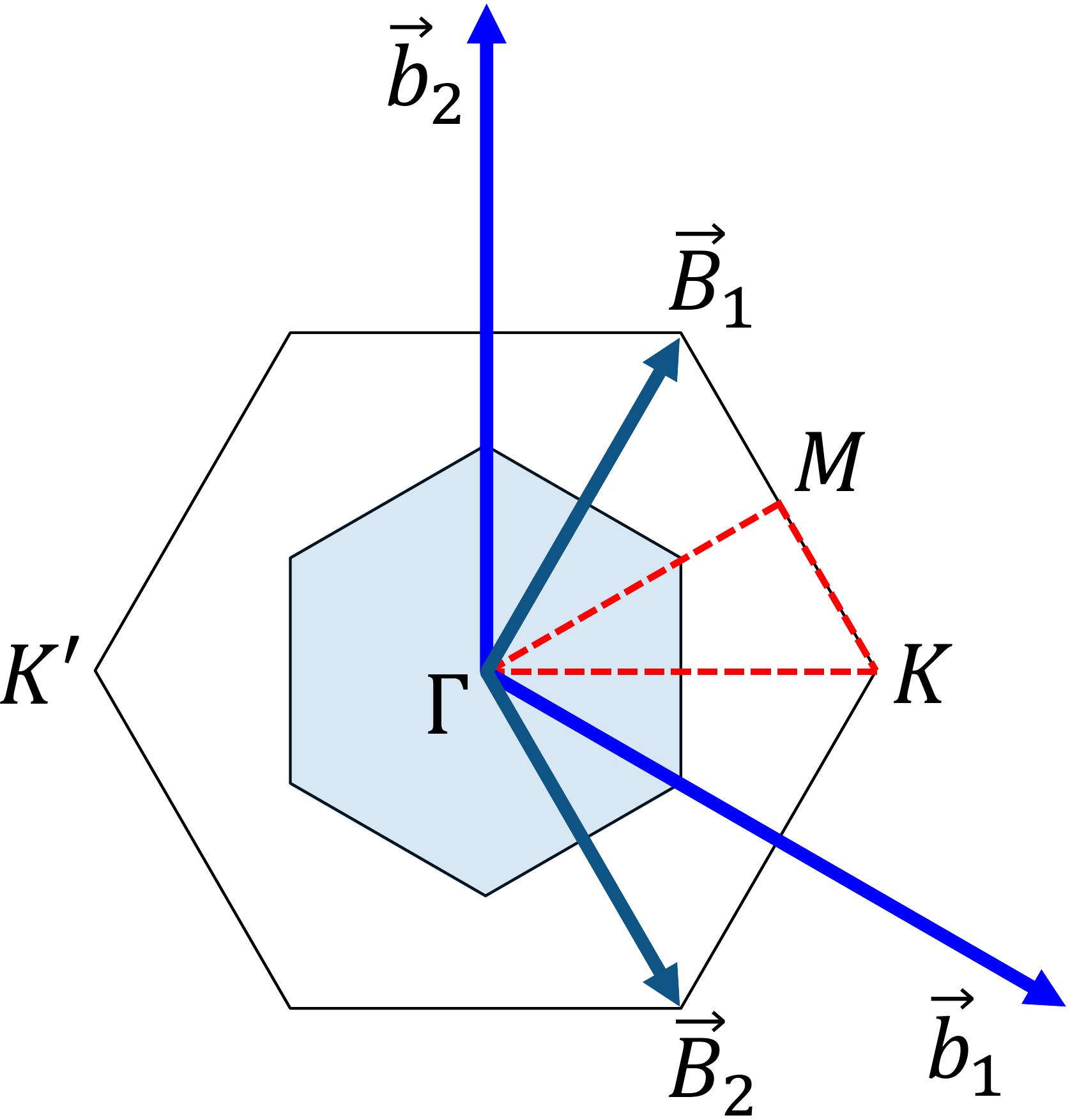}\label{fig:brillouinzone}} \\
	\captionsetup{oneside,margin={0.28cm,0cm}}
	\subfloat[]{\raisebox{2ex}{\includegraphics[width=0.22\textwidth]{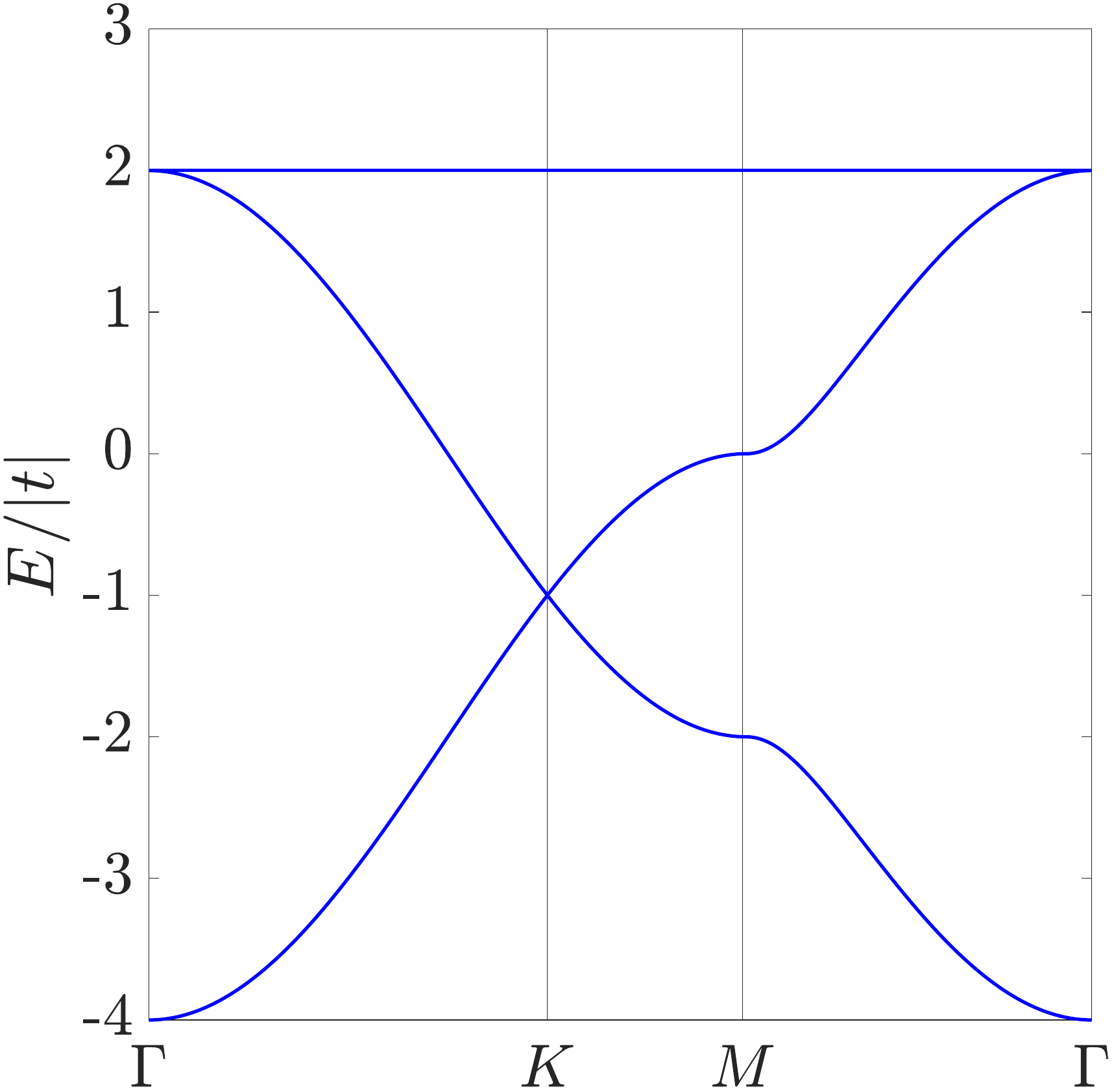}\label{fig:band}}} ~
	\subfloat[]{\raisebox{-0.8ex}{\includegraphics[width=0.24\textwidth]{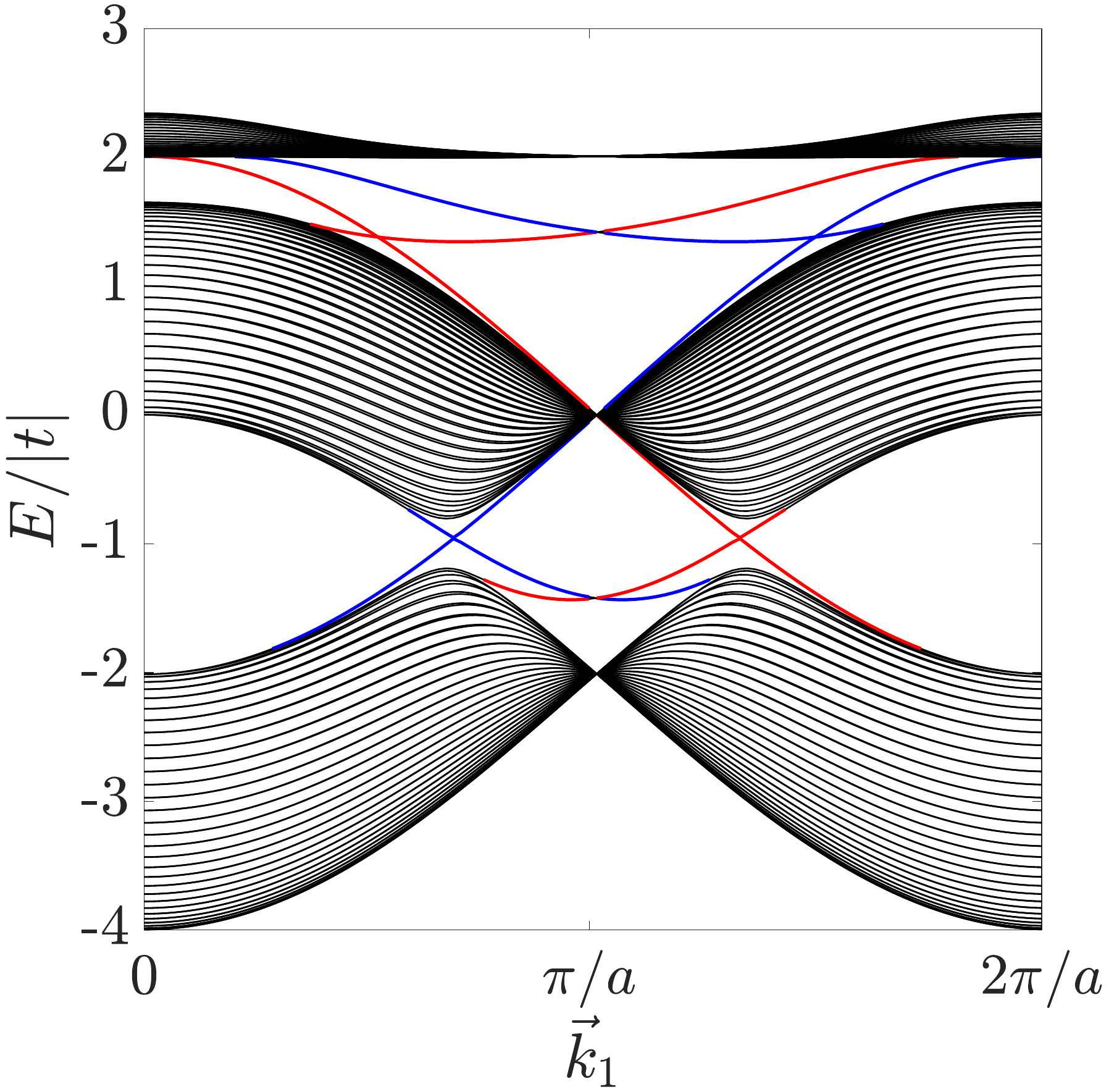}}\label{fig:band_soc}}
	\caption{(a) The kagome lattice structure in real space. Primitive lattice vectors are given by $\vec{a}_1$ and $\vec{a}_2$. $A$, $B$, and $C$ represent three different sublattices. The encircling arrow in the hexagon indicates the sign of intrinsic spin-orbit coupling: $\nu_{ij}= +1$ for anti-clockwise direction and $\nu_{ij}= -1$ for clockwise. Shaded region shows $\sqrt{3}\times\sqrt{3}$ enlarged unit cell, and corresponding lattice vectors are given by $\vec{A}_1$ and $\vec{A}_2$ in figure (b). First and reduced Brillouin zone of kagome lattice is shown in figure (c). $\vec{b}_i$ and $\vec{B}_i$ satisfy $\vec{a}_i\cdot\vec{b}_j = 2\pi\delta_{ij}$ and $\vec{A}_i\cdot\vec{B}_j = 2\pi\delta_{ij}$ respectively. $K$ and $K'$ are two different Brillouin zone corners connected by time reversal operation. (d) Tight binding model band structure of the kagome lattice for $t=1$ in the absence of spin-orbit coupling. (e) Band structure of non-interacting kagome lattice with gapless edge states for $t = 1$ and $\lambda_{\text{SO}} = 0.1$. Periodic boundary condition is imposed only in $\vec{a}_1$ direction and $a = |\vec{a}_1|$. Red and blue color refer to spin up and down edge modes.}
\end{figure}

We start by introducing the Hubbard model with nearest neighbor (NN) intrinsic SOC on the kagome lattice (Fig. \ref{fig:kagomelattice}). The Hamiltonian can be written as
\begin{align}
\hat{H}&=-t\sum_{\langle i,j \rangle,\alpha}\left(\hat{c}_{i \alpha}^{\dagger}\hat{c}_{j \alpha} + h.c.\right) \\
&+ i\lambda_{\text{SO}}\sum_{\langle i,j \rangle \alpha, \beta}\left(\nu_{ij} \hat{c}_{i \alpha}^{\dagger}\sigma_{\alpha\beta}^z \hat{c}_{j\beta} + h.c. \right) + U\sum_{i}\hat{n}_{i\uparrow}\hat{n}_{i\downarrow}. \nonumber
\end{align}
Here, $\hat{c}_{i\alpha}$($\hat{c}_{i\alpha}^{\dagger}$) is the electron annihilation(creation) operator at site $i$ with spin $\alpha=\uparrow,\downarrow$, $\hat{n}_{i\alpha}=\hat{c}_{i\alpha}^\dagger \hat{c}_{i\alpha}$ is the electron number operator at site $i$ with spin $\alpha$, and $\vec{\sigma} = (\sigma^x,\sigma^y,\sigma^z)$ represent the Pauli matrices. $t$, $\lambda_{SO}$, and $U$ are the NN hopping amplitude, NN intrinsic SOC strength, and onsite electron correlation strength respectively. $\braket{i,j}$ denotes pair of NN sites. $\nu_{ij} = +1(-1)$ depends on whether the electron traverses in the anti-clockwise (clockwise) direction from $i$ to $j$ as shown in Fig. \ref{fig:kagomelattice}.

Fig. \ref{fig:brillouinzone} and Fig. \ref{fig:band} show the first Brillouin zone and the energy dispersion of simple tight binding model ($t = 1$, $\lambda_{\text{SO}} = U = 0$) along the dotted line in the Fig. \ref{fig:brillouinzone}. There are Dirac cones at $K$ and $K'$ points stabilized by time reversal, parity, and global $C_3$ rotation symmetry as discussed above. Furthermore, one can see flat band at $E/|t| = 2$ which is closely related to localized feature of wave function in real space. At $\Gamma$ point, there is a degeneracy between flat band and middle band resulting from real space topological effect\cite{Liu_2014_flat,PhysRevB.78.125104}.
In the presence of SOC, similar to the Kane-Mele model\cite{PhysRevLett.95.146802,PhysRevLett.95.226801}, non-trivial mass gap is introduced at the two Dirac cones inducing quantum spin Hall phase at 1/3 filling as shown in Fig. \ref{fig:band_soc}. Furthermore, a non-trivial mass gap is also opened between the dispersive band and the flat band at 2/3 filling at $\Gamma$ point. Therefore, we can realize TI at both 1/3 and 2/3 filling in the non-interacting kagome lattice with finite SOC.

The interacting Hamiltonian is solved here using mean-field theory, and the mean-field Hamiltonian of interaction term can be written as
\begin{align} \label{eq:HFM1}
\hat{H}_{\text{U}}^{\text{HFM}} &\approx U\sum_i\langle \hat{n}_{i\uparrow} \rangle \hat{n}_{i\downarrow} +  \hat{n}_{i\uparrow} \langle \hat{n}_{i\downarrow}\rangle - \langle \hat{n}_{i\uparrow} \rangle \langle \hat{n}_{i\downarrow} \rangle \\
&- \langle \hat{c}_{i\uparrow}^\dagger \hat{c}_{i\downarrow} \rangle \hat{c}_{i\downarrow}^\dagger \hat{c}_{i\uparrow} -  \hat{c}_{i\uparrow}^\dagger \hat{c}_{i\downarrow} \langle \hat{c}_{i\downarrow}^\dagger \hat{c}_{i\uparrow}\rangle + \langle \hat{c}_{i\uparrow}^\dagger \hat{c}_{i\downarrow} \rangle \langle \hat{c}_{i\downarrow}^\dagger \hat{c}_{i\uparrow} \rangle. \nonumber
\end{align}
To simplify Eq. \ref{eq:HFM1}, we introduce charge operator $\hat{Q}_i=\hat{n}_{i\uparrow} + \hat{n}_{i\downarrow}$ and spin operator $\hat{S}_i^a=\frac{1}{2}\sum_{\alpha,\beta}\hat{c}_{i\alpha}^\dagger \sigma_{\alpha\beta}^a \hat{c}_{i\beta}$, where $i$ denotes site index, $\alpha, \beta = \uparrow, \downarrow$ represent spins, and $a = x,y,z$. In terms of charge and spin operators, we can rewrite Eq. \ref{eq:HFM1} as

\begin{align} \label{eq:hfm}
\hat{H}_{\text{U}}^{\text{HFM}} \approx U\sum_i \left(\frac{1}{2}\rho_i\hat{Q}_i - 2\vec{m}_i\cdot\hat{\vec{S}}_i - \frac{1}{4}\rho_i^2 + \left|\vec{m}_i\right|^2\right)
\end{align}
where $\rho_i = \braket{\hat{Q}_i}$ and $\vec{m}_i = \braket{\hat{\vec{S}}_i}$ are charge and magnetic order parameters at site $i$ respectively. Eq. \ref{eq:hfm} implies that the charge ordering $\rho_i$ increases the energy like Coulomb repulsion, and magnetization $\vec{m}_i$ decreases the energy like magnetic energy. Therefore, we can expect the emergence of various charge and spin ordered phases to minimize $\hat{H}_{\text{U}}^{\text{HFM}}$ when $U \neq 0$.
We solve self-consistency equation of given order parameters $\rho_i = \braket{\hat{n}_{i\uparrow} + \hat{n}_{i\downarrow}}$ and $\vec{m}_i = \frac{1}{2}\left\langle\sum_{\alpha\beta}\hat{c}^{\dagger}_{i\alpha}\vec{\sigma}_{\alpha\beta}\hat{c}_{i\beta}\right\rangle$ at 1/3 filling. To capture translation symmetry broken phases in kagome lattice, we consider \textit{$\sqrt{3}\times\sqrt{3}$ enlarged unit cell} (shaded region of Fig. \ref{fig:enlarged_kagome}).

\begin{figure}
	\centering
	\subfloat{\label{fig:phase_diagram}\includegraphics[width=0.45\textwidth]{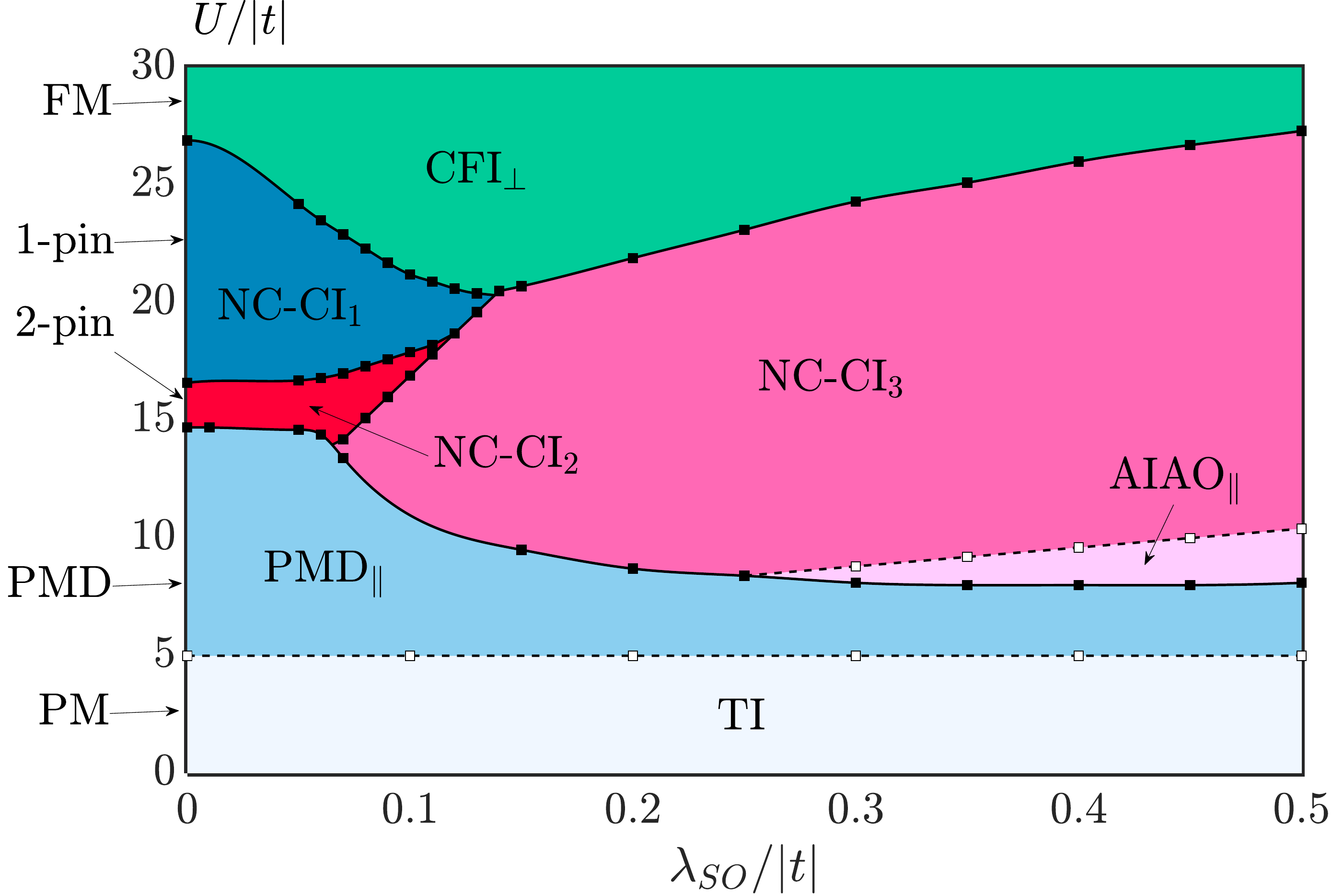}}
	\caption{Phase diagram of 1/3 filling kagome lattice with onsite Hubbard interaction and intrinsic SOC. $x$ and $y$ axis are NN SOC strength $\lambda_{\text{SO}}/|t|$ and onsite Hubbard interaction strength $U/|t|$ respectively. Filled (empty) squares and solid (dotted) lines represent first (second) order phase transition. Each abbreviation in the phase diagram denotes paramagnetic metal (PM), pinned metal droplet (PMD), ferromagnetic metal (FM), 1,2-pinned metallic phase (1,2-pin), topological insulator (TI), coplanar pinned metal droplet (PMD$_\parallel$), perpendicular Chern ferromagnetic insulator (CFI$_\bot$), coplanar all-in-all-out (AIAO$_\parallel$), and non-collinear Chern insulator type-1,2,3 (NC-CI$_\text{1,2,3}$).}
\end{figure}

\begin{figure}
	\centering
	\subfloat[PMD/PMD$_\parallel$]{\raisebox{2ex}{\includegraphics[width=0.22\textwidth]{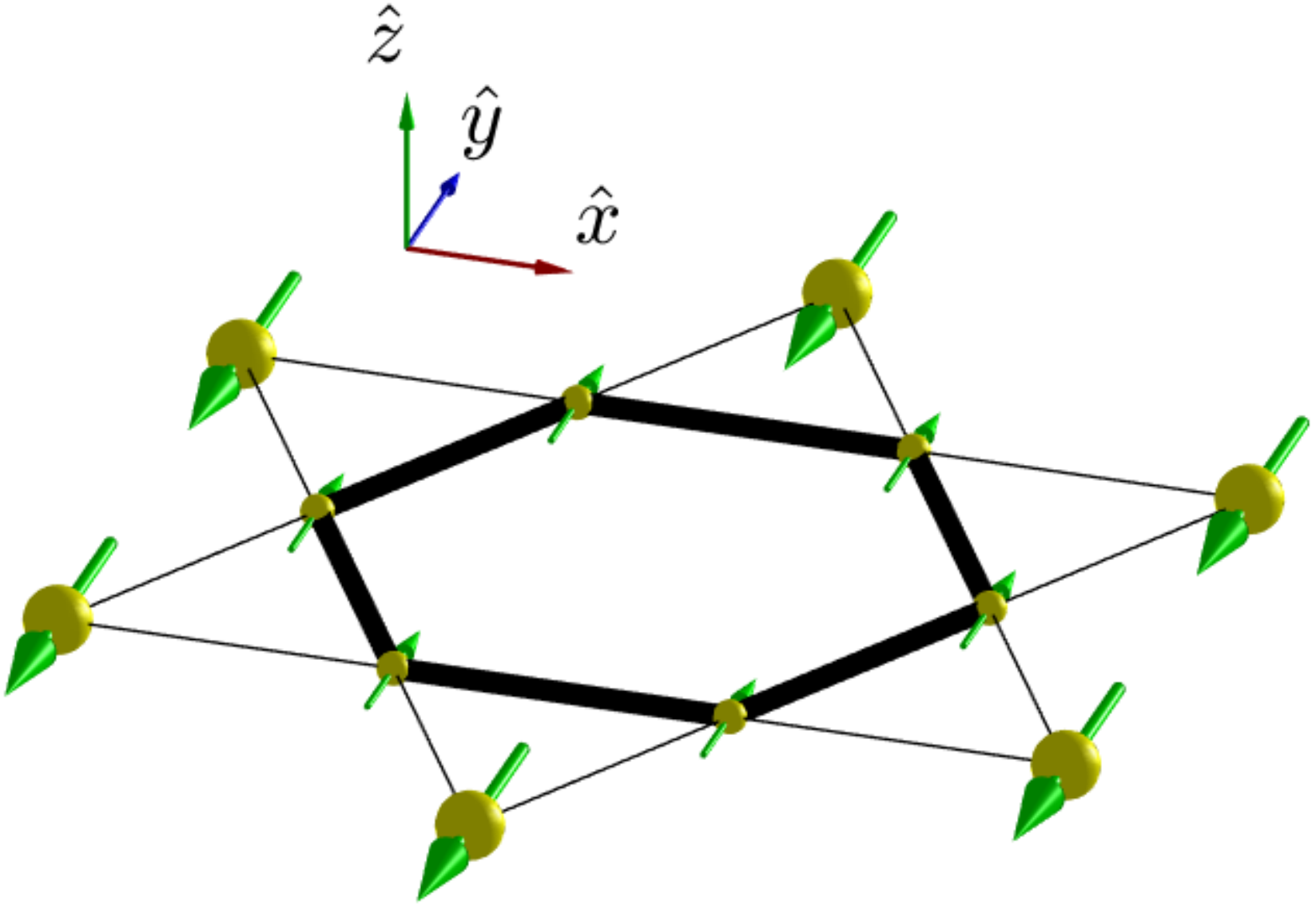}\label{fig:PMD}}} \qquad
	\subfloat[FM/CFI$_\bot$]{\raisebox{0.5ex}{\includegraphics[width=0.2\textwidth]{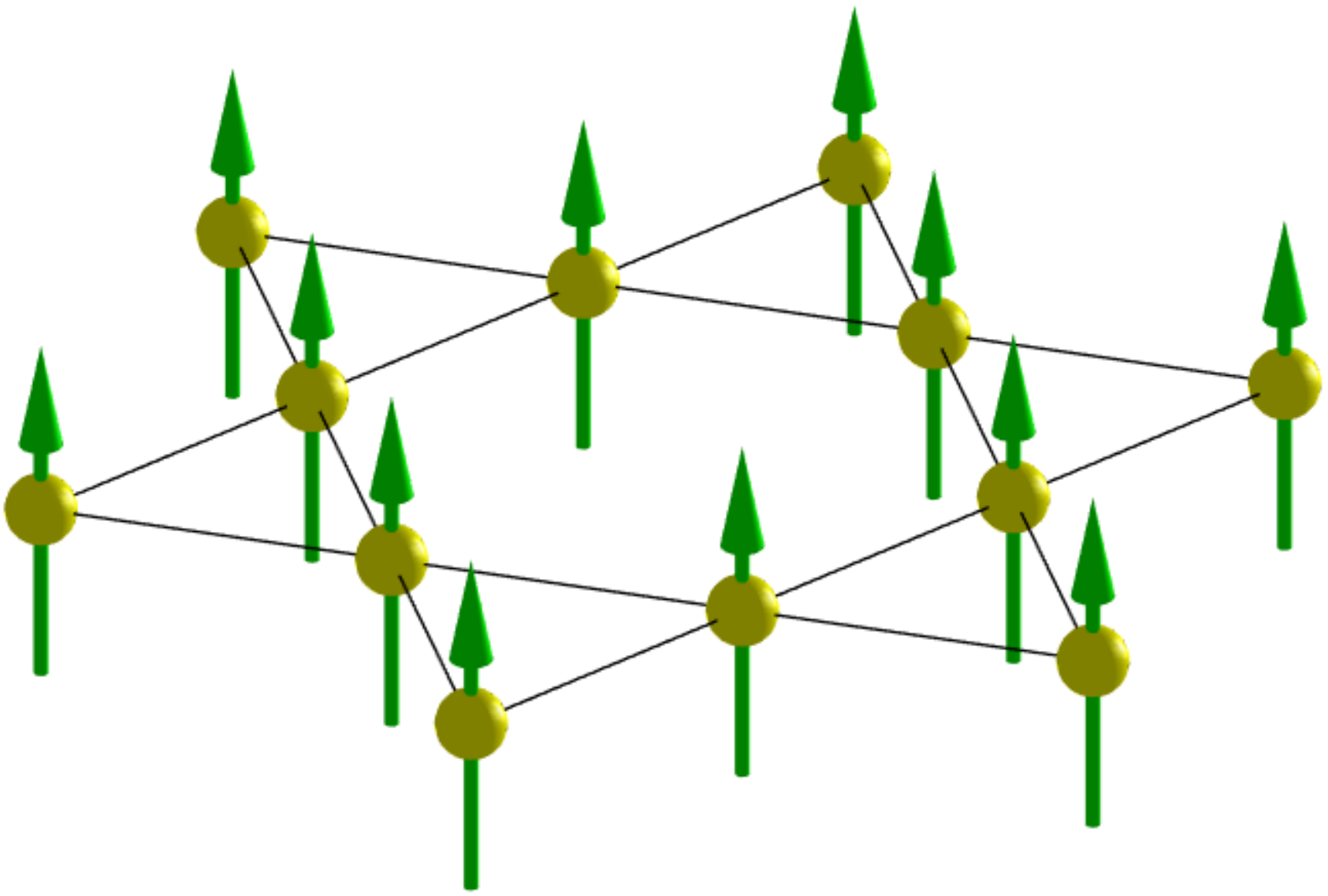}\label{fig:FM}}} \\
	\subfloat[1-pin]{\raisebox{1ex}{\includegraphics[width=0.21\textwidth]{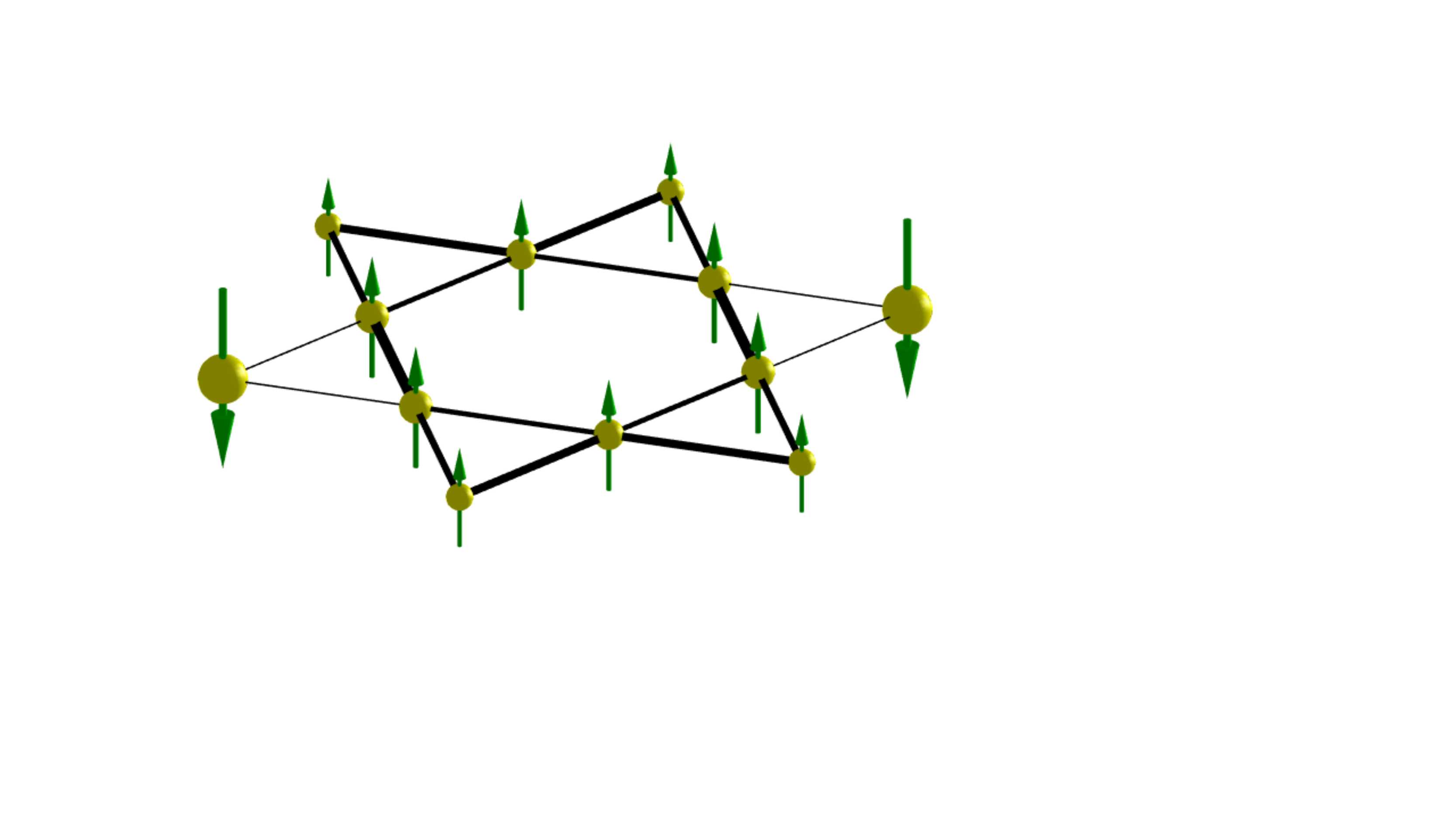}}\label{fig:1pin}} \qquad
	\subfloat[2-pin]{\raisebox{1ex}{\includegraphics[width=0.21\textwidth]{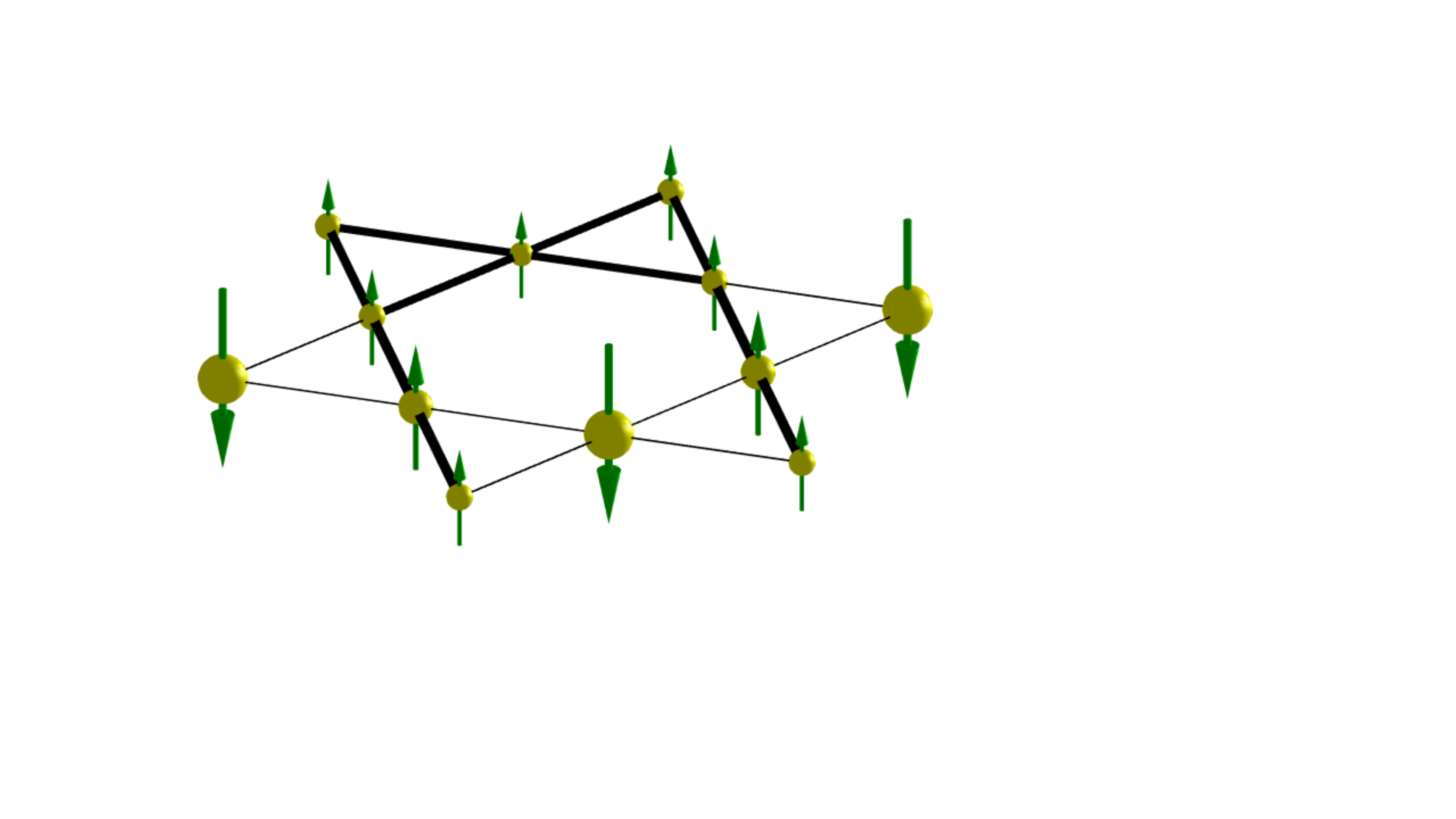}}\label{fig:2pin}} \\
	\subfloat[AIAO$_\parallel$]{\raisebox{1ex}{\includegraphics[width=0.26\textwidth]{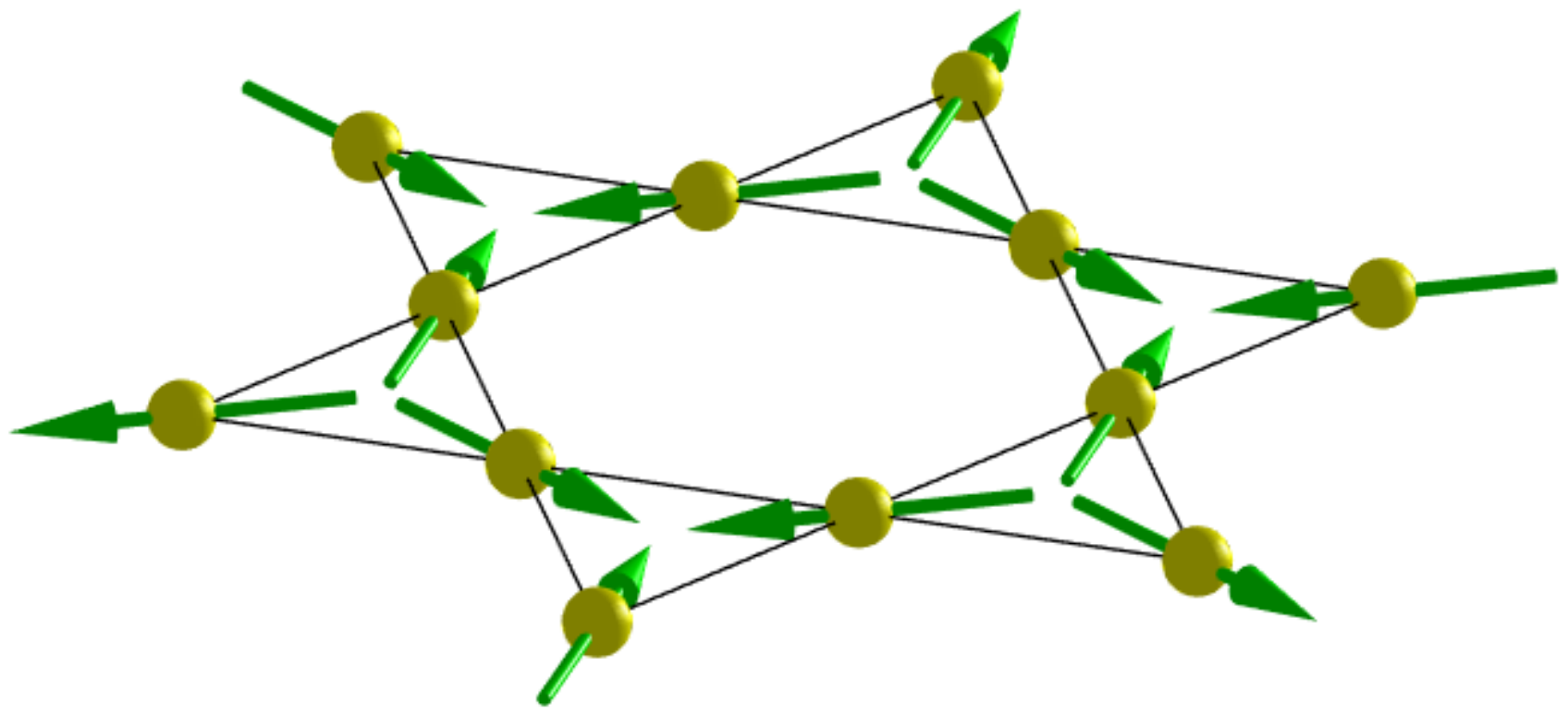}}\label{fig:AIAO}} ~
	\subfloat[NC-CI$_3$]{\raisebox{1ex}{\includegraphics[width=0.22\textwidth]{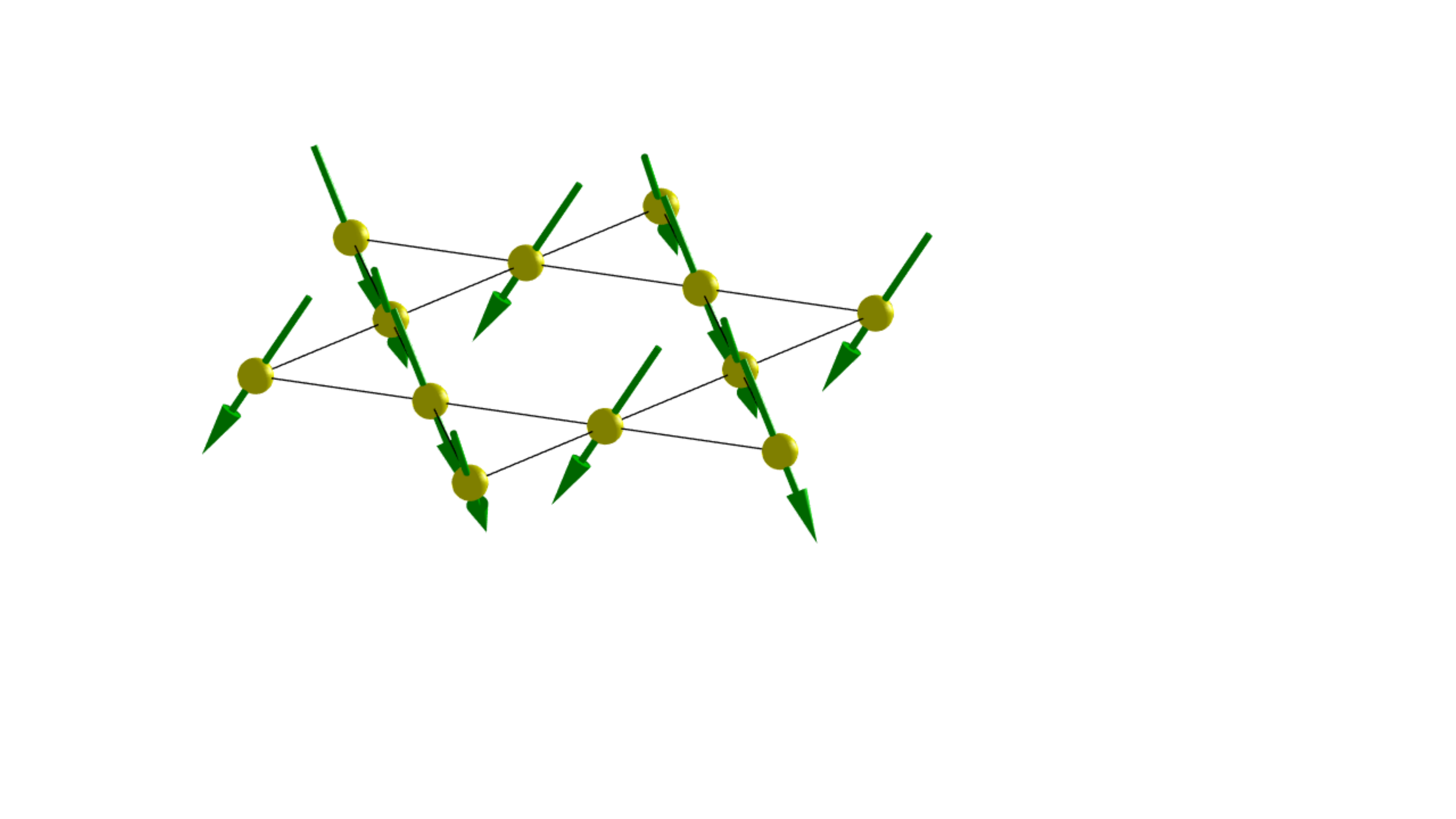}}\label{fig:CI3}}
	\caption{Charge and spin ordering of various phases: (a) Pinned metal droplet/Coplanar pinned metal droplet (PMD/PMD$_\parallel$) (b) Ferromagnetic metal/Perpendicular Chern ferromagnetic insulator (FM/CFI$_\bot$) (c) 1-pinned metallic phase (1-pin) (d) 2-pinned metallic phase (2-pin) (e) Coplanar all-in-all-out (AIAO$_\parallel$) (f) Non-collinear Chern insulator type-3 (NC-CI$_\text{3}$). Thickness of lines, size of circles, and length of arrows represent relative magnitude of bond ordering, charge ordering $\rho_i$, and spin ordering $\vec{m}_i$ respectively in each figure.}
\end{figure}

Let's first discuss the case when $\lambda_{\text{SO}}=0$. From now on, we set $t = 1$ and other parameters are normalized by $t$. As mentioned above, non-interacting kagome lattice is a semi-metal at 1/3 filling with band touching at the Dirac points. In the presence of interactions, unique charge and spin orderings are being developed as shown in Fig. \ref{fig:phase_diagram}. Near $U_c \sim 5$, the second order phase transition occurs and the system is divided into two opposite spin sectors. One sector is pinned at the corner of the enlarged kagome unit cell, and the other sector makes metallic ring at each hexagonal plaquette as shown in Fig. \ref{fig:PMD}. Therefore we denote this phase as pinned metal droplets (PMD) and it has been reported in Ref. \onlinecite{PhysRevB.89.155141}. For extremely large $U \sim 27$, the charge ordering is evenly distributed with equal magnetization aligned in the same direction on each site leading to a ferromagnetic ordering with $\rho_i = 2/3$ and $|\vec{m}_i| = 1/3$ (Fig. \ref{fig:FM}). Due to spin polarization, the up spin and down spin sectors are well separated. Hence, at 1/3 filling, the system is a ferromagnetic metal (FM) with the valence and conduction band touching at $\Gamma$ point. At large $U$, the dominant term of the mean-field Hamiltonian, $\braket{\hat{H}_{\text{U}}^{\text{HFM}}} = U\sum_i\left(\frac{1}{4}\rho_i^2 - |\vec{m}_i|^2\right)$ is zero for ferromagnetic ordering, hence, being the ground state. The two phases mentioned above have already been reported previously\cite{PhysRevB.82.075125,PhysRevB.89.155141}. 

\begin{figure}
	\centering
	\subfloat{\label{fig:effective_energy}\includegraphics[scale=0.29]{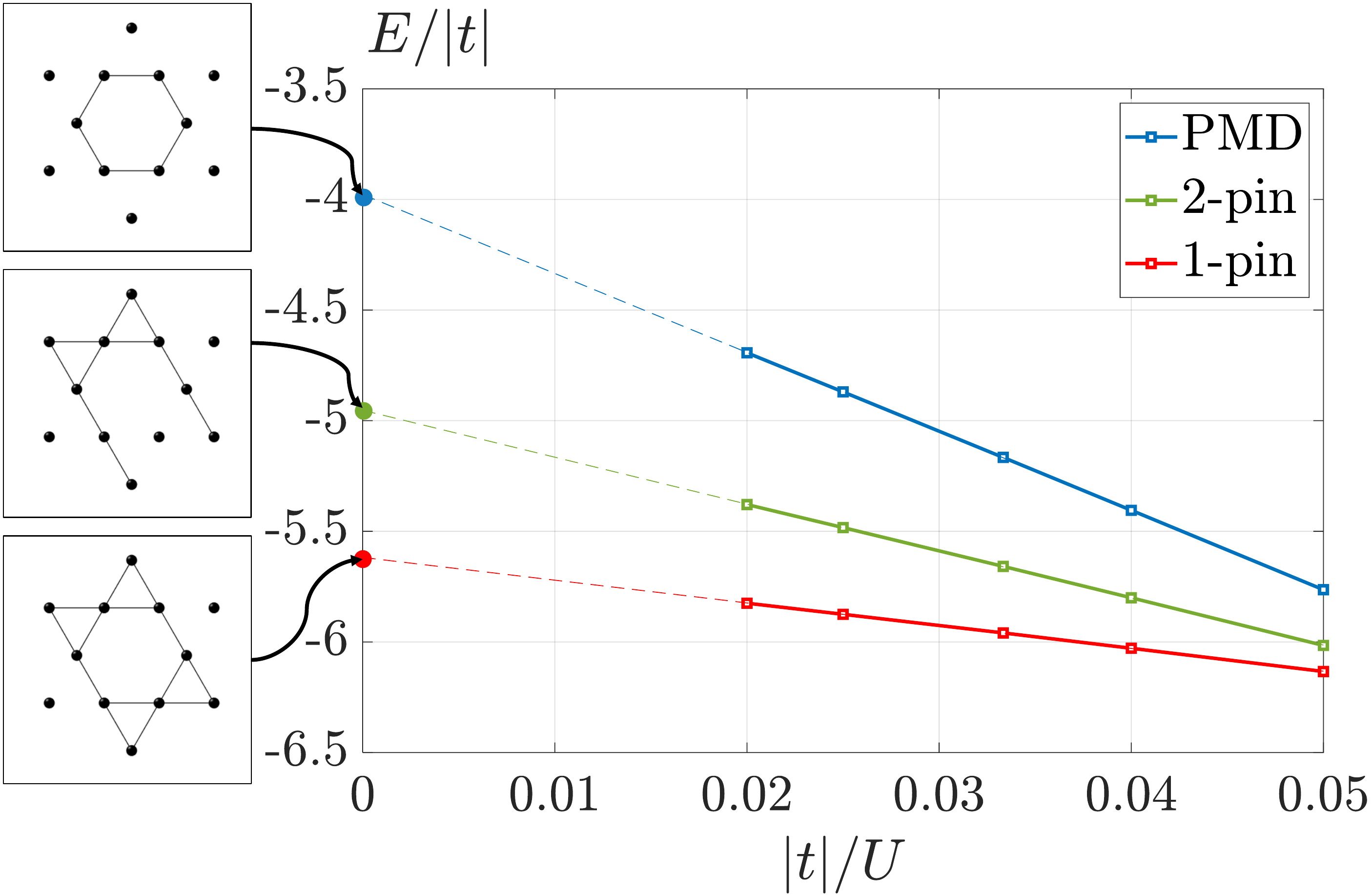}}
	\caption{Graph of pinned metal droplet, 2-pinned metallic, and 1-pinned metallic phase energy with varying $U/|t|$ value. Left figures show effective tight binding models for each phase at $U/|t| \to \infty$.}
\end{figure}

In between PMD and FM, however, we discover new interesting magnetic metallic phases so called \textit{1-(2-) pinned metallic} phase. These two phases, as the names indicate, have 1 and 2 pinned (localized) electrons in the enlarged unit cell as shown in Fig. \ref{fig:1pin} and Fig. \ref{fig:2pin} respectively. Similar to PMD, two opposite spin sectors are divided into localized and itinerant electron parts. Distinct with PMD, however, these phases are formed by coupled conducting chains resulting in metals. For each pinned phase, the Coulomb energy of pinned site and hopping attached to it converge to 0 as $U$ increased. This is because the system avoids doubly occupied states to minimize the electron interaction. Therefore, we can effectively treat each $n$-pin ($n = 0,1,2,3$) phase as simple tight binding model with eliminating hopping attached to pinned sites (Fig. \ref{fig:effective_energy}). Here, 3-pin phase corresponds to PMD phase. According to this effective tight binding model, the system tends to have less pinned sites to lower the energy as $U$ increased. This is why these 1-pin and 2-pin metallic phases are stabilized in the middle of PMD and FM.

\begin{figure}
	\centering
	\captionsetup{oneside,margin={0.28cm,0cm}}
	\subfloat[PMD$_\parallel$]{\label{fig:PMDXY_band}\includegraphics[width = 0.24\textwidth]{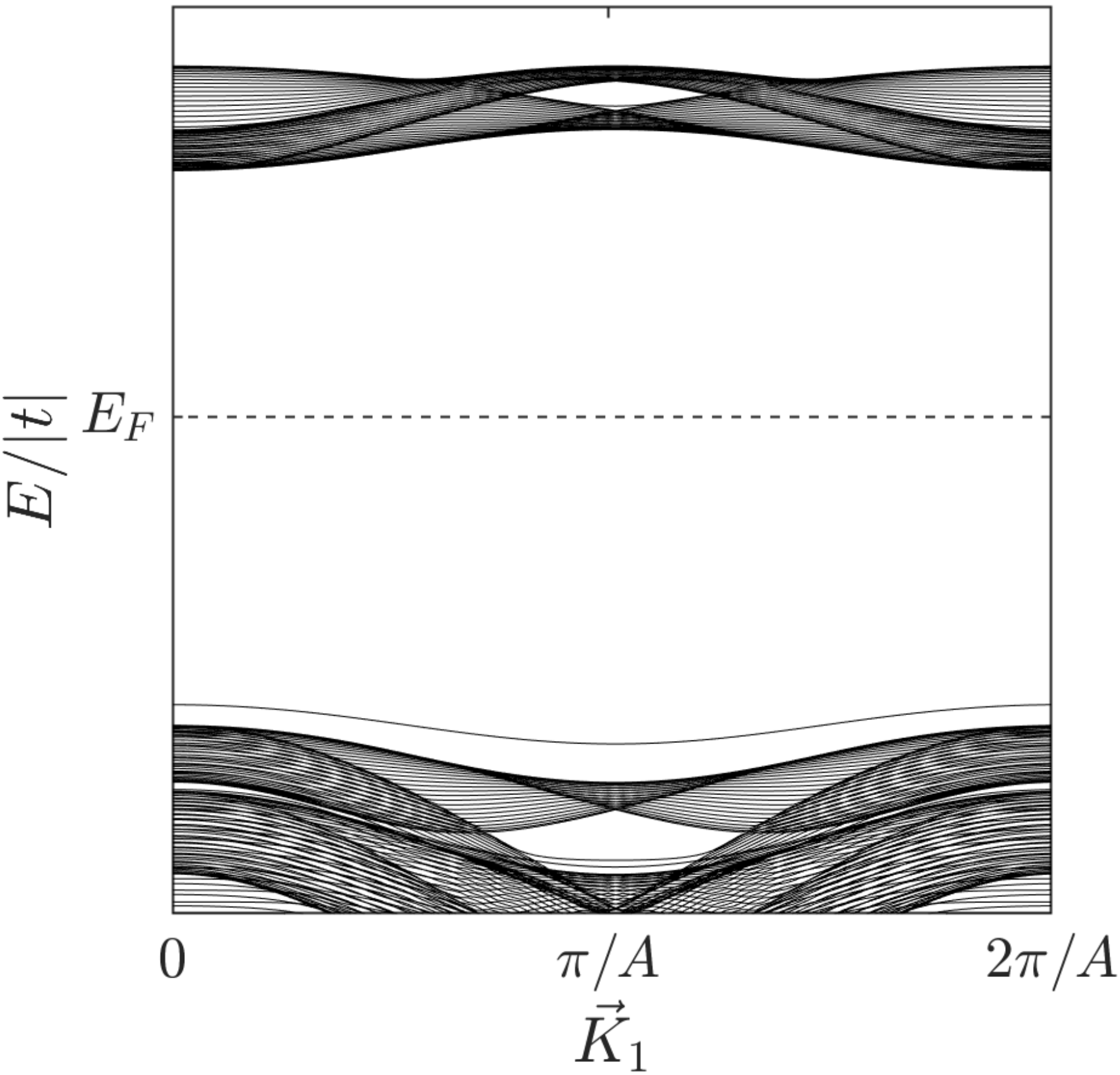}}
	\subfloat[CFI$_\bot$]{\label{fig:CFI_band}\includegraphics[width = 0.24\textwidth]{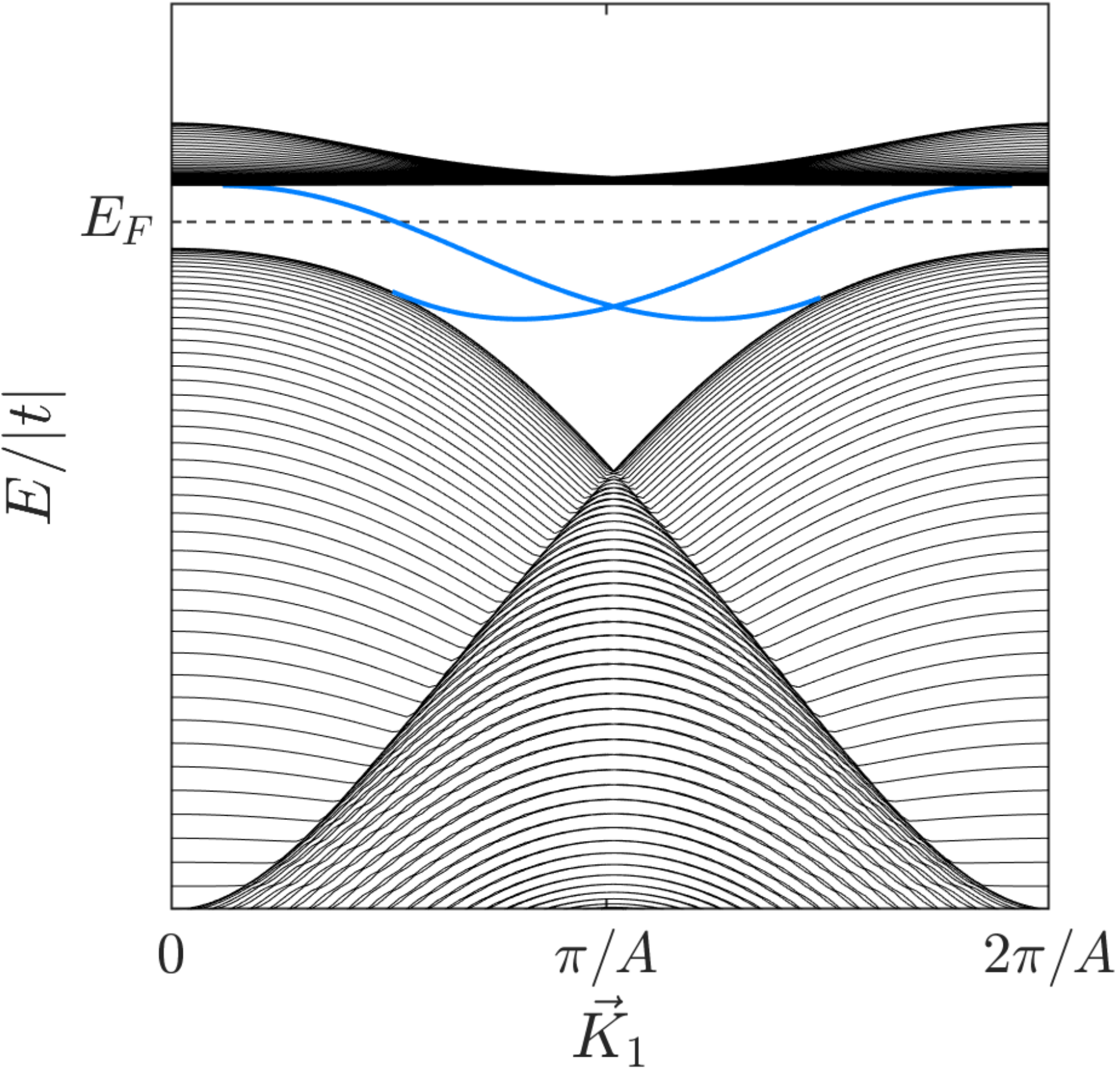}} \\ \subfloat[AIAO$_\parallel$]{\label{fig:AIAO_band}\includegraphics[width = 0.24\textwidth]{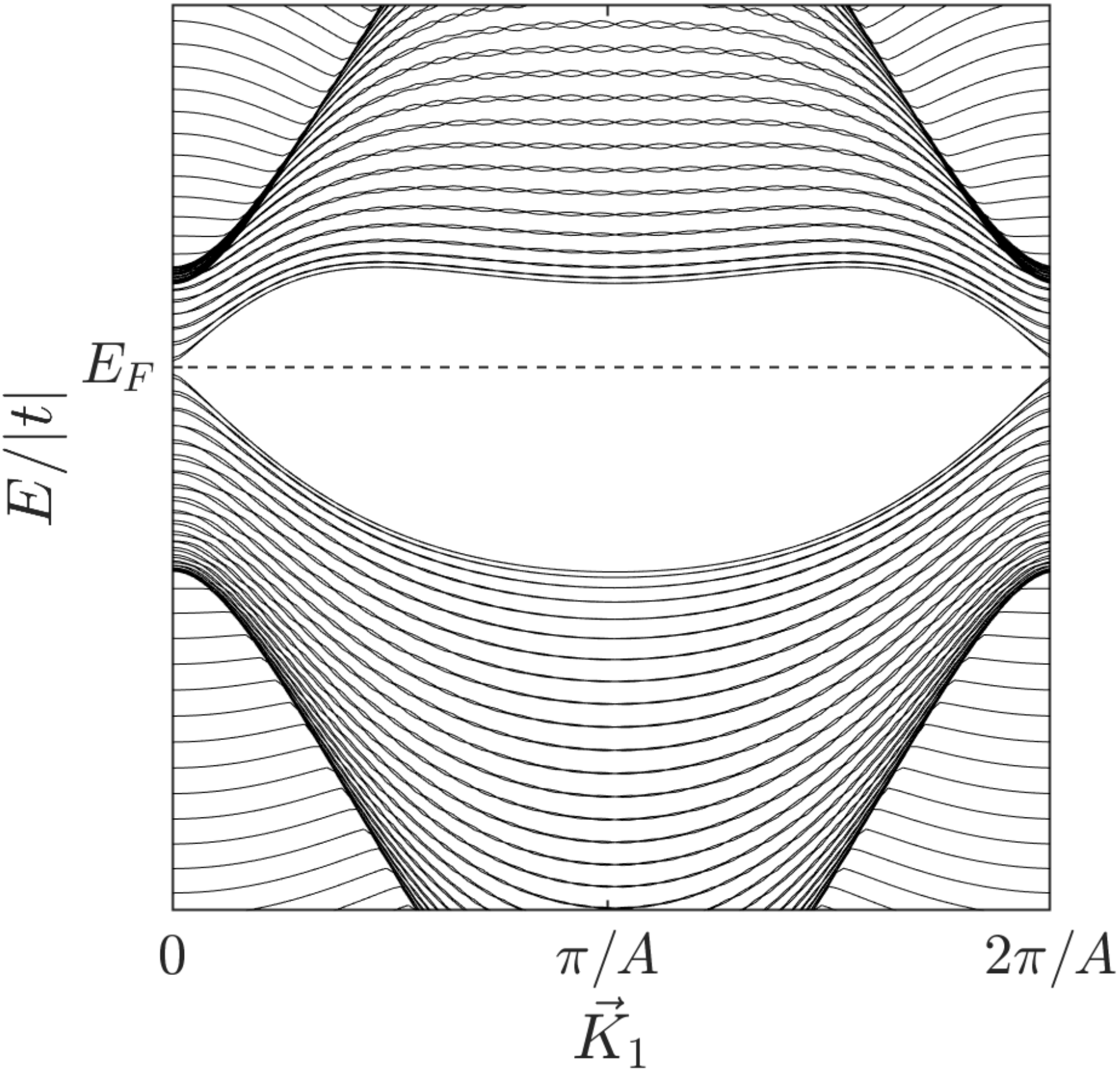}}
	\subfloat[NC-CI$_\text{3}$]{\label{fig:CI3_band}\includegraphics[width = 0.24\textwidth]{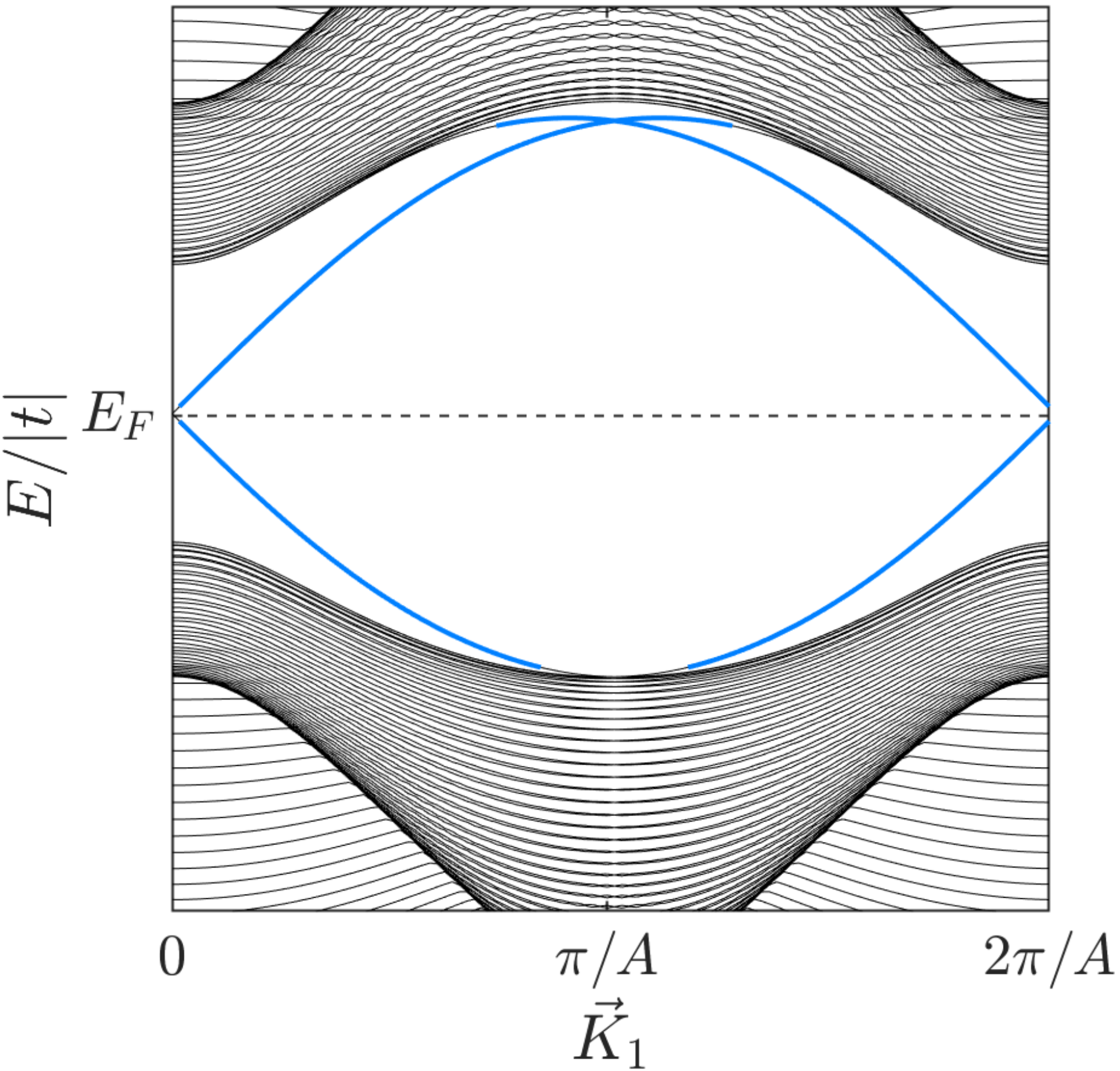}} \\
	\subfloat[NC-CI$_\text{1}$]{\label{fig:CI1_band}\includegraphics[width = 0.24\textwidth]{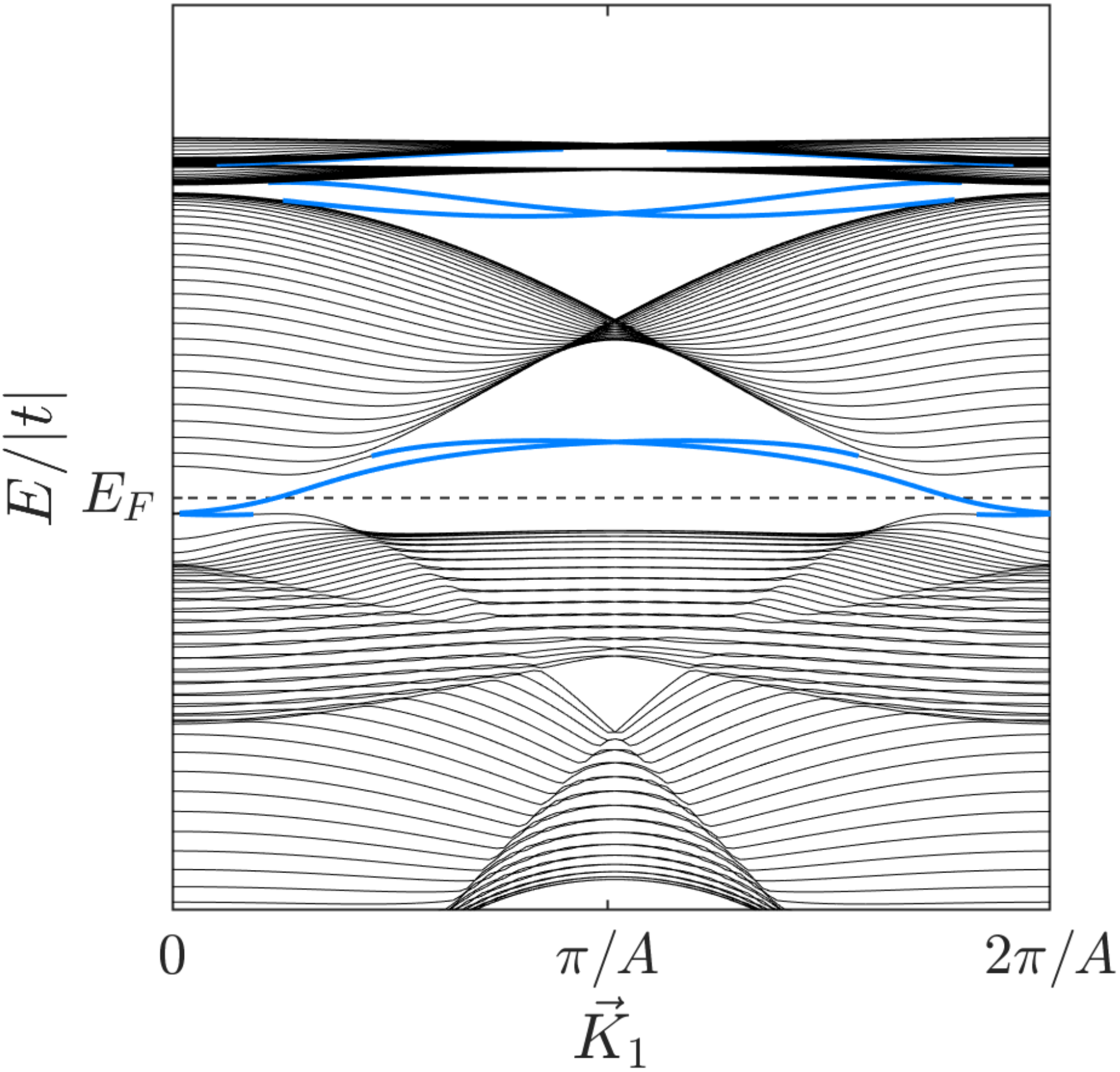}}
	\subfloat[NC-CI$_\text{2}$]{\label{fig:CI2_band}\includegraphics[width = 0.24\textwidth]{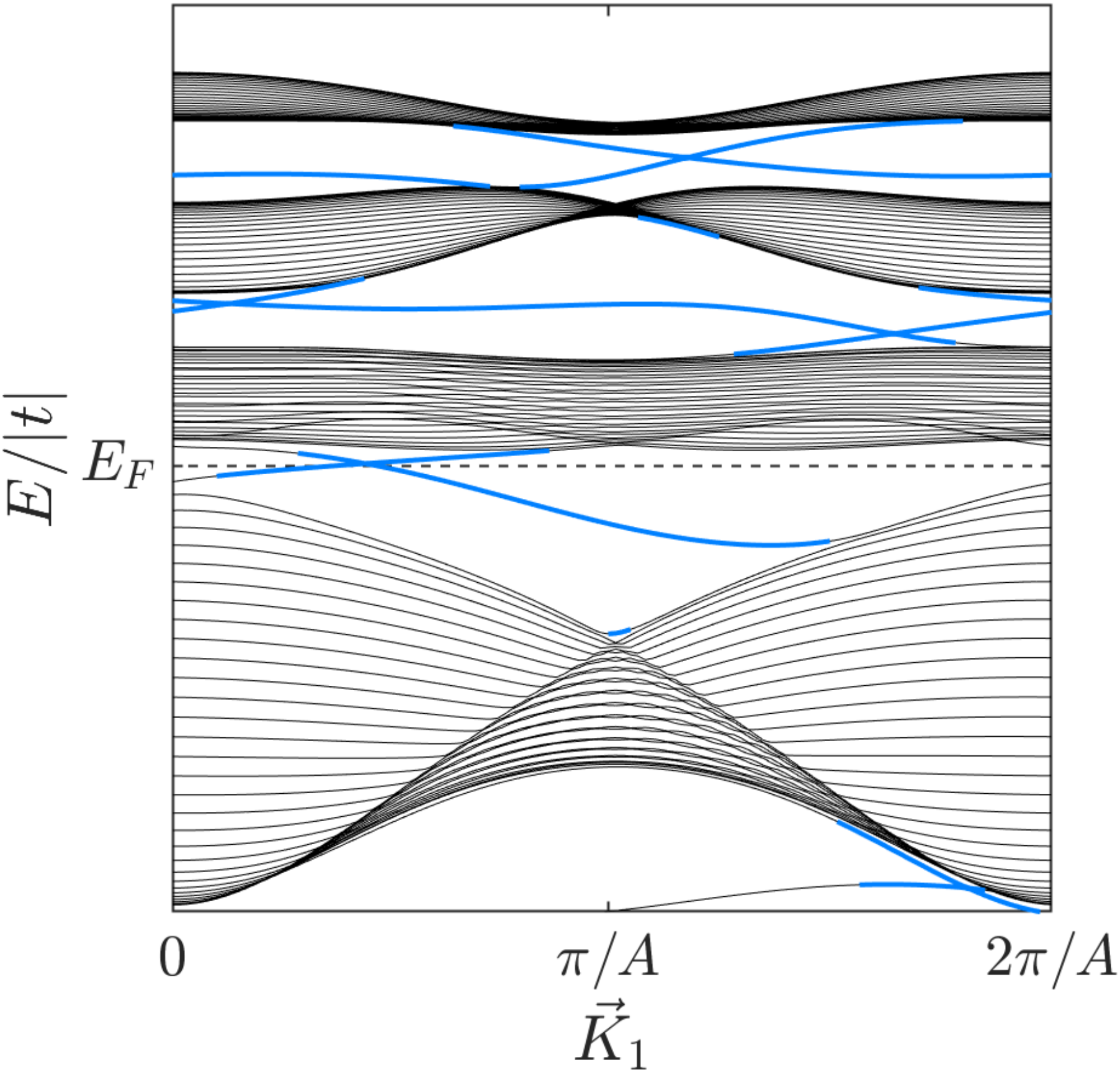}}
	\caption{Band structure with stripe geometry near the Fermi level of (a) Coplanar pinned metal droplet (PMD$_\parallel$) (b) Perpendicular Chern ferromagnetic insulator (CFI$_\bot$) (c) Coplanar all-in-all-out (AIAO$_\parallel$) (d) Non-collinear Chern insulator type-3 (NC-CI$_\text{3}$) (e) Non-collinear Chern insulator type-1 (NC-CI$_\text{1}$) (f) Non-collinear Chern insulator type-2 (NC-CI$_\text{2}$). Periodic boundary condition is imposed in only $\vec{A}_1$ direction and $A = |\vec{A}_1|$ (see Fig. \ref{fig:enlarged_kagome}). In each figure, blue lines represent edge states and the dotted line is the position of Fermi level $E_F$.}
\end{figure}

Now we discuss the case when SOC is present. In the presence of SOC, the system becomes TI at 1/3 filling for $U < U_c$ as discussed before. Increasing $U>U_c$, there is a second order phase transition to PMD. However, the magnetization $\vec{m}_i$ prefers parallel to the kagome plane than perpendicular direction, thus labeled as PMD$_\parallel$. This is because the SOC breaks $SU(2)$ symmetry and hybridizes spin up and down sector. Also, magnetization breaks time reversal symmetry spontaneously thus edge states are not topologically protected. Regarding to effective model of PMD and previous discussion, this phase becomes trivial insulator as shown in Fig. \ref{fig:PMDXY_band}. For very large $U$ limit, the spins are aligned ferromagnetically and the magnetization prefers to align perpendicular to kagome plane. SOC opens a gap between the flat band and the second dispersive band as shown in Fig. \ref{fig:CFI_band} inducing a Chern insulator phase. Hence, we denote this phase as Chern ferromagnetic insulator (CFI$_\bot$), where subscript $\bot$ indicates perpendicular magnetization to the kagome plane.

PMD$_\parallel$ and CFI$_\bot$ emerge even for small SOC and can be understood by $SU(2)$ symmetry breaking from PMD and FM. However, sufficiently large SOC ($\lambda_{\text{SO}} \gtrsim 0.25$) opens a new horizon of magnetism. Since SOC acts like a hopping parameter with hybridizing two spin sectors, first order phase transition occurs from PMD$_\parallel$ and the system chooses another kind of coplanar magnetic phase beyond critical $U$ value. Unlike PMD$_\parallel$, there is no pinned site and spin orderings are aligned exactly at $120\degree$ to each other with all-in all-out configuration in the kagome unit cell as shown in Fig. \ref{fig:AIAO}. We denote this phase as coplanar all-in all-out (AIAO$_\parallel$).

Between AIAO$_\parallel$ and CFI$_\bot$, a new type of spin ordering appears. As shown in Fig. \ref{fig:CI3}, the spins are aligned exactly in AIAO$_\parallel$ fashion in the kagome plane, at the same time, having additional component in the perpendicular direction.
 Therefore spin ordering looks like a mixing of AIAO$_\parallel$ and CFI$_\bot$. This phase has non-zero Chern number and we denote this phase as non-collinear Chern insulator type-3 (NC-CI$_\text{3}$). Fig. \ref{fig:CI3_band} shows the band structure of this phase along with the edge states at the Fermi surface. The origin of non-trivial Chern number of bands and non-collinear magnetization are related in the following way\cite{PhysRevB.62.R6065,PhysRevB.74.085105}. Let's define scalar spin chirality for each triangle on the kagome lattice as $\chi_{\triangle} = \vec{m}_i \cdot \vec{m}_j \times \vec{m}_k$, where $i,j,k$ are corners of triangle in anti-clockwise direction order. Since the total scalar spin chirality on the kagome lattice should be zero, we can deduce scalar spin chirality on the hexagonal plaquette $\chi_{\hexagon}$. This scalar spin chirality operates as gauge flux in the kagome lattice and yields spin Berry phase due to the part of mean-field Hubbard term $\sum_i\vec{m}_i\cdot\hat{\vec{S}}_i$. This gauge effect opens a non-trivial mass gap between bands with non-zero Chern number. 

Apart from NC-CI$_\text{3}$, we also find other kinds of Chern insulator with non-zero scalar spin chirality. If we turn on the SOC from the 1-pin and 2-pin phase, two opposite spin sectors are mixed with spin anisotropy, so that spin ordering becomes non-collinear way. Similar to the NC-CI$_\text{3}$, each triangle and hexagonal plaquette of kagome lattice has finite scalar spin chirality, $\chi_{\triangle}, \chi_{\hexagon} \neq 0$, but different magnetization pattern from NC-CI$_\text{3}$. Therefore, we denote this phase as Chern insulator with non-collinear spin ordering type-1 and type-2 respectively (NC-CI$_\text{1,2}$). Fig. \ref{fig:CI1_band} and Fig. \ref{fig:CI2_band} show the edge states at Fermi level.

In conclusion, we have studied interplay of electron correlation and SOC in the kagome lattice  at 1/3 filling based on Hartree-Fock mean-field theory and effective model analysis. In strongly interacting limit, it turns out that kagome lattice geometry results in several new types of phases with finite SOC.  The interaction induces unique charge and spin ordered phases like PMD, 1-pin, 2-pin, and FM. Emerging 1-pin and 2-pin metallic phases between PMD and FM can be understood by effective tight binding model analysis. In addition, introducing SOC leads to unique magnetic metallic phase AIAO$_\parallel$, topologically trivial insulator phase PMD$_{\parallel}$, and topologically non-trivial insulating phases such as CFI$_\bot$, and NC-CI$_{\text{1,2,3}}$. Especially, PMD$_\parallel$, CFI$_\bot$, and NC-CI$_{1,2}$ can be explained by $SU(2)$ symmetry breaking with SOC from PMD, FM, 1-pin, and 2-pin respectively. Finally, for a wide range of phase diagram, the NC-CI type phases appear having a finite spin chirality and non-trivial spin Berry phase due to interplay of SOC and electron interaction. As an extension of our work, it will be interesting to consider exchange effect of multi-layer kagome lattice on the magnetization and topological properties, which we leave as future work.

\bibliography{bibfile}

\end{document}